\documentclass{article}

\usepackage{PRIMEarxiv}
\usepackage[table,xcdraw]{xcolor}
\usepackage[utf8]{inputenc} 
\usepackage[T1]{fontenc}    
\usepackage{hyperref}       
\usepackage{url}            
\usepackage{booktabs}       
\usepackage{amsfonts}       
\usepackage{nicefrac}       
\usepackage{microtype}      
\usepackage{lipsum}
\usepackage{fancyhdr}       
\usepackage{graphicx}       
\graphicspath{{media/}}     
\usepackage[affil-it]{authblk}
\usepackage{adjustbox}
\usepackage{amsmath}

\title{Machine Learning Based Lens-Free Shadow Imaging 
Technique for Field-Portable Cytometry
\thanks{\textit{\underline{Citation}}: 
\textbf{Vaghashiya, R.; Shin, S.; Chauhan, V.; Kapadiya, K.; Sanghavi, S.; Seo, S.; Roy, M. Machine Learning Based Lens-Free Shadow Imaging Technique for Field-Portable Cytometry. Biosensors 2022, 12, 144. https://doi.org/10.3390/bios12030144}} 
}

\author[1,4]{Rajkumar Vaghashiya}
\author[2,4]{Sanghoon Shin}
\author[1]{Varun Chauhan}
\author[1]{Kaushal Kapadiya}
\author[1]{Smit Sanghavi}
\author[2,5]{Sungkyu Seo}
\author[3,5]{Mohendra Roy} 

\affil[1]{%
  Department of Computer Engineering, Pandit Deendayal Energy University, Gandhinagar, India}
\affil[2]{%
  Department of Electronics and Information Engineering, Korea University, Sejong 30019, Korea}
\affil[3]{%
  Department of Information and Communication Technology, Pandit Deendayal Energy University, Gandhinagar 38207, India
  }
\affil[4]{%
  \small \textnormal{These authors contributed equally to this study}}
\affil[5]{%
  \small \textnormal{Corresponding authors: mohendra.roy@gmail.com, mohendra.roy@ieee.org, sseo@korea.ac.kr}}

\begin{document}
\maketitle

\begin{abstract}
Lens-free Shadow Imaging Technique (LSIT) is a well-established technique for the characterization of microparticles and biological cells. Due to its simplicity and cost-effectiveness, various low-cost solutions have been evolved, such as automatic analysis of complete blood count (CBC), cell viability, 2D cell morphology, 3D cell tomography, etc. The developed auto characterization algorithm so far for this custom-developed LSIT cytometer was based on the hand-crafted features of the cell diffraction patterns from the LSIT cytometer, that were determined from our empirical findings on thousands of samples of individual cell types, which limit the system in terms of induction of a new cell type for auto classification or characterization. Further, its performance is suffering from poor image (cell diffraction pattern) signatures due to its small signal or background noise. In this work, we address these issues by leveraging the artificial intelligence-powered auto signal enhancing scheme such as denoising autoencoder and adaptive cell characterization technique based on the transfer of learning in deep neural networks. The performance of our proposed method shows an increase in accuracy $>$98\% along with the signal enhancement of $>$5 dB for most of the cell types, such as Red Blood Cell (RBC) and White Blood Cell (WBC). Furthermore, the model is adaptive to learn new type of samples within a few learning iterations and able to successfully classify the newly introduced sample along with the existing other sample types.
\end{abstract}


\section{Introduction}

\begin{figure}
  \centering
  \includegraphics[width=\textwidth,keepaspectratio]{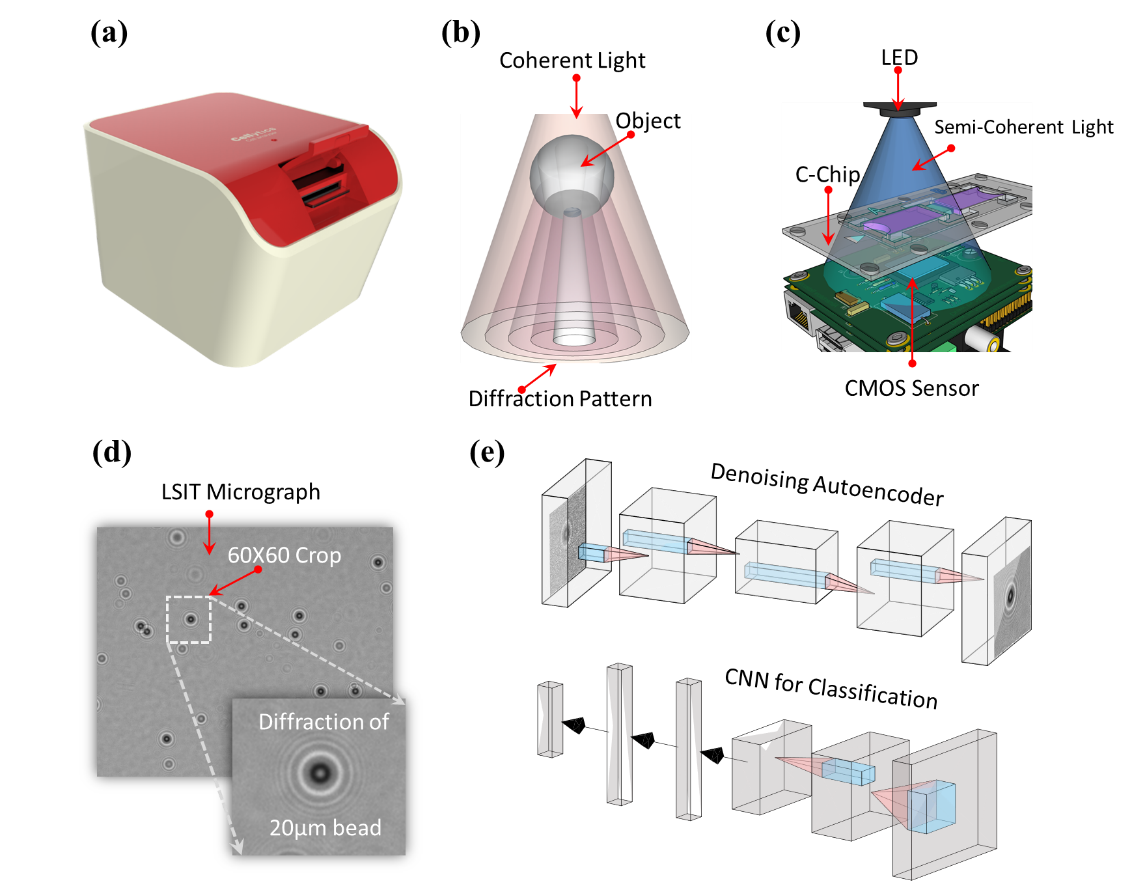}
  \caption{Schematics of the LIST setup and the proposed neural network architecture for the auto characterization of LSIT micrographs. (a) LSIT platform (Cellytics) (b) schematic of the principle of diffraction, i.e., shadow, pattern generation of a micro-object, (c) schematic of the LSIT imaging setup showing the simplicity of the setup, (d) schematic of the dataset creation process by automatic cropping individual cell diffraction pattern from the whole LSIT micrograph, (e) the schematic of the proposed denoising and classification architecture. Here, the denoising autoencoder enhances the signal of the individual cells which is then fed to the CNN module for classification.}
  \label{fig:Fig1}
\end{figure}

Lens-free Shadow Imaging Technique (LSIT) is a well-established technique for the characterization of microparticles and biological cells \cite{Mudanyali2010}. This technique is widely popular for its simple imaging structure and cost-effectiveness. It comprises a lens-less detector such as a complementary metal-oxide semiconductor (CMOS) image sensor, a semi-coherent light source, such as light-emitting diode (LED), and a disposable cell chip (C-Chip). The absence of lens or other optical arrangements allows it to fit into a very small space, thereby reducing the size of the overall system as described in Figure \ref{fig:Fig1}a, the LSIT platform (Cellytics) built within a dimension of 100 × 120 × 80 mm\textsuperscript{3}. Since this arrangement consists of a few components, most of which are easily available and at a low price, therefore it reduces the overall cost of the system \cite{Roy2015}. This simple and cost-effective nature facilitates the feasibility of the LSIT for applications in the field of point-of-care systems or telemedicine systems \cite{Roy2014,Roy2016,Roy2015a}.

Recent advancements in machine learning, especially deep learning, have facilitated many applications concerning medical diagnostics \cite{Liu2017,Yu2018,Liu2019,Ardila2019,Im2018,Wang2018,Ballard2020}, and have been widely adopted in the field of microscopy \cite{Rivenson2017,Ozcan2020,Wang2020}. Especially, deep learning has been incorporated with the LSIT \cite{Ozcan2020}, where it is has been used to enhance the resolution of the LSIT micrographs \cite{DeHaan2020} and enabled polarization-based holographic microscopy \cite{Liu2020}.

In our previous work, we have successfully developed the LSIT imaging system for the complete blood count using an analytical model based on handcrafted features \cite{Roy2014} that can automatically segment out the individual cells from a whole frame LSIT micrograph and subsequently analyze them based on the handcrafted parameters. However, the performance of the system is dependent on the uniform illumination as well as the strong signatures of the microparticle samples. Since the diffraction signature of a microparticle depends on the size as well as the signal-to-noise ratio of the particle, therefore any background noise can affect the overall performance of the auto characterization system. Further, the handcrafted approach of finding the features for every additional cell line is time-consuming and prone to subjective errors. To address these limitations, in this work, we have developed an artificial intelligence (AI) powered signal enhancement scheme for the LSIT micrographs that can enhance the signal quality (signal to noise ratio (SNR)) for various cell lines in a heterogeneous cell sample. For this, we employed the autoencoder-based denoising scheme \cite{Kunapuli2019}. Further, we have developed an auto characterization method based on a convolutional neural network \cite{Indolia2018,Yamashita2018} (CNN) architecture to classify the various cell lines from the LSIT micrograph. Here, we have first introduced the transfer of learning scheme in a neural network, which can leverage the feasibility to introduce new cell types to the algorithm and thus learn their characteristics within a few iterations. Thus, the LSIT platform saves time and computation resources required to learn to classify the additional cell types along with the existing ones.

In this article, we have described the detailed methods adopted for the designing as well as optimization of various parameters to design a suitable model with better accuracy. These optimized models are simple and light-weight, and require a smaller number of samples for effectively learning the cell signatures. The details are as given in the following sections.

\section{Methods}

\subsection{LSIT Imaging Setup}

The schematic of our proposed setup (Figure \ref{fig:Fig1}a) is as shown in Fig. \ref{fig:Fig1}. When light from the coherent or semi-coherent source passes through a micro-object, it produces characteristic diffraction, i.e., a shadow, pattern of the object \cite{Seo2009,Mudanyali2010a} as shown in Fig. \ref{fig:Fig1}b. These diffraction patterns are prominent just beneath the sample, typically a few hundred micrometers away from the sample plane, from where they are captured using a high-density image sensor such as CCD or CMOS \cite{Seo2009} (Fig. \ref{fig:Fig1}c). As these signatures are significant enough to be captured by the bared image sensor, it does not require any kind of lens arrangement \cite{Jin2012}. In our proposed setup, we used a pinhole conjugated semi-coherent LED light source with a peak wavelength of 470 ± 5 nm (HT-P318FCHU-ZZZZ, Harvatek, Hsinchu, Taiwan). The diffraction patterns were captured using a 5-megapixel CMOS image sensor (EO-5012M, Edmund Optics, Barrington, NJ, USA), and a custom-developed C-Chip (Infino, Seoul, Korea) was used to hold the cell samples \cite{Roy2015,Roy2016,Roy2015a}. All these components can fit in a compact dimension of 100 mm × 120 mm × 80 mm. Due to the absence of a lens-based setup, the field-of-view of this system is about 20 times that of a conventional optical microscope at 100×. This high-throughput nature provides an extra advantage to characterize several thousand cells within a single digital frame.

\subsection{Preparation of Various Cell Lines}

In this work we used various cell lines, starting from red blood cell (RBC), white blood cell (WBC), cancer cell lines HepG2 (human liver cell-line) and MCF7 (human breast cancer cell-line), and polystyrene microbeads of 10 µm and 20 µm. The preparations of these cell lines are as follows \cite{Roy2015,Roy2014}. The use of human whole blood in the experiment was approved by the Institutional Review Board (Approval No. \# 2021AN0040 of Korea University Anam Hospital (Seoul, Korea).

\textbf{RBC:} The RBC samples were prepared from the whole blood samples that were collected from the Korea University Anam Hospital under IRB approval. The samples were diluted about 16,000 times by using RPMI solution (Thermo Scientific, Waltham, MA, USA) \cite{Roy2015,Roy2014}.

\textbf{WBC:} First, Ficoll solution (Ficoll-Paque\textsuperscript{TM} Plus, GE Healthcare, Chicago, IL, USA) was used to isolate mononuclear cells from the whole blood. The samples of peripheral blood mononuclear cells (PBMCs) obtained using the Ficoll solution are mixtures of lymphocytes and monocytes. To separate these two cell types, the MACS (Magnetic-activated cell sorting) device and antibodies (Miltenyi Biotec, Bergisch Gladbach, Germany) were utilized. The helper-T cells in the lymphocytes were separated using the CD4 antibody (\#130-090-877), and the cytotoxic-T cells with the CD8 antibody (\# 130-090-878). Finally, 10 µL of this solution was then loaded into the unruled C-Chip cell counting chamber \cite{Roy2015,Roy2014}.

\textbf{HepG2:} The HepG2 cell lines were prepared from the American Type Culture Collection (ATCC HB-8065) and incubated in a high-glucose medium (DMEM, Merck, Darm-stadt, Germany) with 10\% heat-inactivated fetal bovine serum, 0.1\% gentamycin, and a 1 penicillin\/streptomycin solution under 95\% relative humidity and 5\% CO\textsubscript{2} at 370°C. The developed cells were then trypsinized and separated from 24 well pate and incubated from 2–5 min at 370°C. These cells were then diluted with DMEM solution \cite{Roy2015,Roy2014}.

\textbf{MCF7:} The MCF7 cell samples were prepared from the American Type Culture Collection (ATCC HTB-22). The cells were preserved in a solution of DMEM containing 1\% pen-icillin/streptomycin solution, 0.1\% gentamycin, and 10\% calf serum at 95\% relative humidity and 5\% CO2 at 370°C. These cells were then trypsinized and separated from the 24 well pate. These separated cells were then incubated for 2–5 min at 370°C. The cells were then washed with DMEM solution. 10 µL of this solution was then loaded in the C-Chip \cite{Roy2015,Roy2014}.

\textbf{Polystyrene microbead:} The 10 µm and 20 µm bead samples were prepared by diluting the respective polystyrene microbeads (Thermo Scientific, Waltham, MA, USA) with de-ionized water \cite{Roy2015,Roy2014}.

\subsection{Dataset Creation}

A whole frame LSIT image (of cell diffraction patterns) contains an average of  $\sim$500 diffraction patterns of microparticles. Deep learning-based architectures utilize the features of each class, and typically require a minimum of a few hundred diffraction patterns of each cell type for optimal learning. Therefore, we cropped individual diffraction patterns of each cell type (that were verified using a traditional microscope) with a window of 66×66 pixels as shown in Fig. \ref{fig:Fig1}d. This window size included the complete sample signature along with a minimal background that would provide complete cell-line information during the auto-feature selection process in learning algorithms. We further augment this base sample set by rotating the individual diffraction patterns with an increasing angle of 10 degrees clockwise. Finally, a dataset of 1980 samples for each of the 6 cell lines and microparticles was created, totaling 11,880 samples for all the classes under study. The typical architecture of a CNN is illustrated in Fig. \ref{fig:Fig1}e. As many learning algorithms are black-box models, it is difficult to ascertain the optimal cell-signal size that covers the majority of the information and minimal background. Naturally, a smaller cell-size would need lesser computation and have lesser noise. Hence, we created the dataset for 60×60, 56×56, 50×50, 46×46, 40×40, and 36×36 cell sizes as input sets, with each set further divided into training and test folds. As the data augmentation used sample rotation, the splitting of the dataset into the train, validation, and test folds needs to be done while keeping a check on data leakage. Augmented samples distributed across the train and test sets may bias the model and may give a wrong estimate of its performance as the test data may not be of entirely “unseen” samples. Accounting for this, the 1980 samples of each class were carefully split into 1490 training samples, 166 validation samples, and 324 testing samples.

\begin{figure}[h]
  \centering
  \includegraphics[width=\textwidth,keepaspectratio]{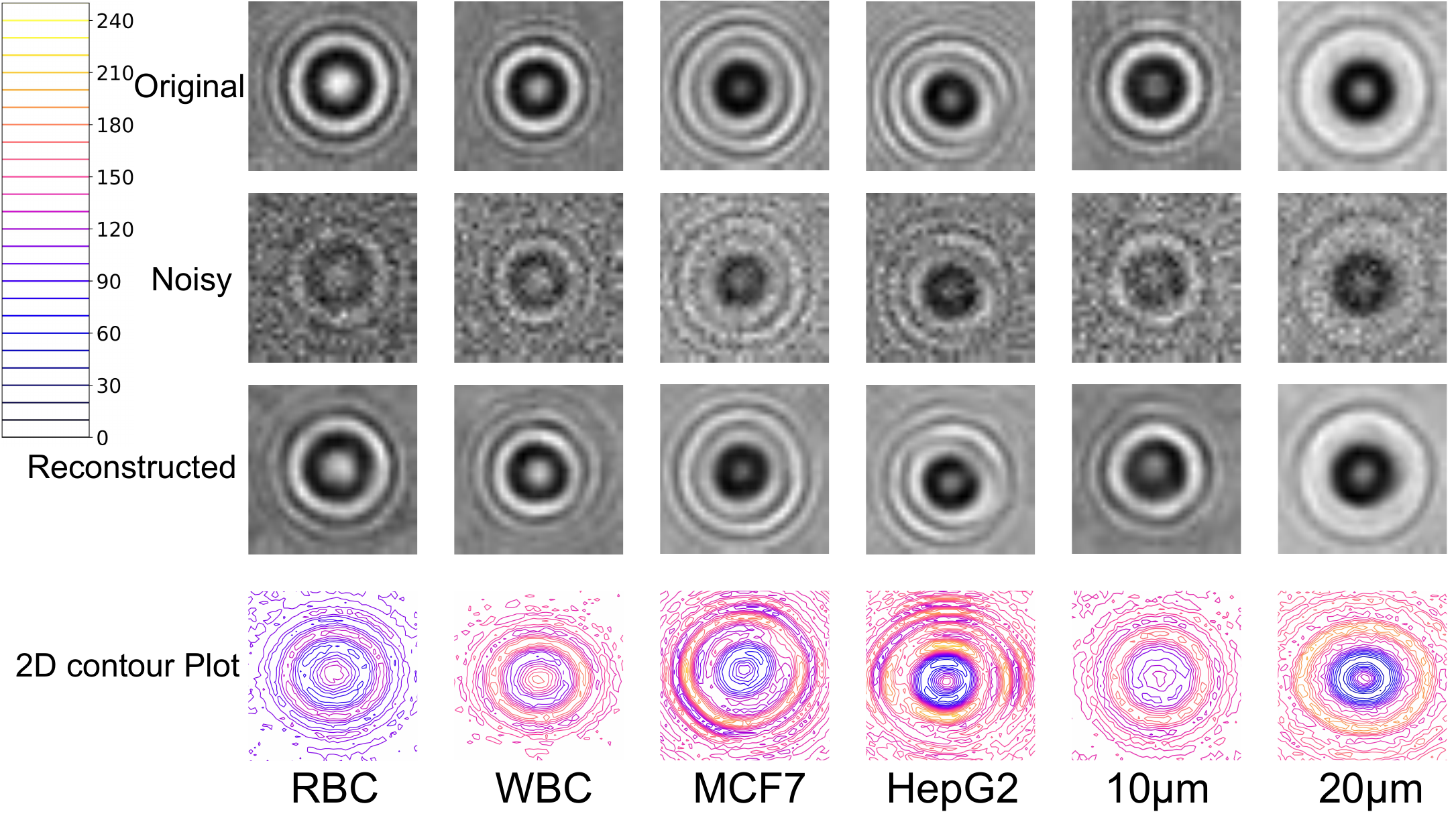}
  \caption{Reconstructed results from the optimized CNN. The top row is the original LSIT image (cell diffraction pattern) of a single RBC, WBC, MCF7, HepG2, 10 µm beam, and 20 µm bead. The second row is the noisy version (with variance 100) of the corresponding original images (cell diffraction pattern). The third row is the denoised version of the corresponding original images (cell diffraction pattern) from the noisy image (cell diffraction pattern). The fourth row is the 2D intensity contour plot of the original image (cell diffraction pattern) to show the unique signature of each of these cell lines.}
  \label{fig:Fig2}
\end{figure}

Though the cell-lines may seem visually similar, there are significant differences in the statistical distributions of the pixel illumination intensity in the cell diffraction pattern as revealed in our exploratory data analysis. The 2D contour plots (in Figure \ref{fig:Fig2}) show the observed variances. Hence, it is possible for intelligent algorithms to automatically identify and utilize the descriptive features in signal enhancement as well as classification.

\subsection{Denoising Modality}

For denoising of the LSIT micrographs, we adopted the concept of autoencoder \cite{Vincent2010}. An autoencoder is an unsupervised scheme that scuffles to recreate the input at its output. It consists of an input layer (x), an output layer (r), and a hidden layer (h). The hidden layer h termed as a code layer stands for the input in a revised dimension. The whole net-work structure can be labelled into two parts. The first part is an encoder, which tries to code the input as \(h = f(x)\), and the second part is a decoder which tries to recreate the input from the reduced code layer as \(r = g(h)\), where r is the recreated assortment of input x. Basically, it tries to attain \(r = g(f(x))\). However, this is not a linear transformation since the model is enforced to learn the significant features of the input to encode it into the code layer.

In this work, we specifically used the denoising version of the autoencoder. Traditionally, the autoencoders try to reduce the loss as \[L(x, g(f(x)))\]. However, the denoising autoencoder attempts to reduce the cost as \[L(x, g(f(x\prime)))\] where x’ is the noisy form of the input x. And we tried two different methods to design the denoising architectures, namely, extreme learning machine (ELM) and convolutional neural network (CNN).

\subsubsection{ELM}

This is a single hidden layer fully connected architecture \cite{Cambria2013}. In this method, the input weights are initiated randomly and kept intact. Only the output weights take part in the learning process through a straightforward learning method \cite{Cambria2013,Bucurica2015,Huang2015}. For N arbitrary input samples $x_i \in R^n$ and their counterpart targets $t_i \in R^m$, the ELM achieves this mapping using the following relation as shown in Equation (\ref{eq:1}).

\begin{equation} \label{eq:1}
  H \beta = T  
\end{equation}

Here, H is the hidden layer output matrix, $\beta$ is the output weight matrix, i.e., between the hidden layer and the output layer, and T is the target matrix or matrix of desired out-put \cite{Bucurica2015}. From Equation (\ref{eq:1}), we can obtain the $\beta$ using Moore–Penrose pseudoinverse \cite{Cambria2013} as shown in Equation (\ref{eq:2}).

\begin{equation} \label{eq:2}
    \beta = (H^T H)^{-1} H^T T    
\end{equation}

In the extended sequential learning form of ELM, the $\beta$ can update sequentially. This provides an added advantage of updating the learning whenever a new type of sample is available, thus providing the flexibility of transfer of learning. The $\beta$ update mechanism \cite{Liang2006,VanSchaik2015} is as shown in Equation (\ref{eq:3}).

\begin{equation} \label{eq:3}
    \beta_n = \beta_{n-1} + P_n^{-1} ( T_n-H_n \beta_{n-1} ) H_n^T    
\end{equation}

Here

\begin{equation} \label{eq:4}
    P_n = P_{n-1} + H_n H_n^T    
\end{equation}

For n = 1,

\begin{equation} \label{eq:5}
    P_{n-1} = P_0 = (\frac{1}{C}+H_0 H_0^T )    
\end{equation}

Here $H_0$ is the hidden layer output with the 1st sample or 1st batch of samples \cite{Roy2019}.

\subsubsection{CNN}

Convolutional neural networks (CNN) \cite{Liu2017,Indolia2018,Yamashita2018} are a type of neural network widely used in the analysis of spatial data such as image classification and object segmentation. In this network, two-dimensional kernels are used to extract the spatial features from the input patterns, using a convolution operation between the kernel and the input. The typical architecture of a CNN is as shown in Fig. \ref{fig:Fig1}e. Here, the kernel is shared spatially by the input or by the feature map. The feature at the location \((i,j)\) in the \(k^{th}\) feature map of the \(l^{th}\) layer can be evaluated as shown in Equation (\ref{eq:6}).

\begin{equation} \label{eq:6}
    Z_{i,j,k}^l= (W_K^l )^T X_{i,j}^l + b_k^l    
\end{equation}

Here, \(W_K^l\) and \(b_k^l\) are the weights and the bias vector of the \(k^{th}\) filter in the \(l^{th}\) layer. Here the weight layer is shared spatially which reduces the complexity. \(X_{i,j}^l\) is the value of the input at location \((i,j)\) of the \(l^{th}\) layer. The nonlinearity in this network can be obtained by introducing the activation function, denoted here as \(g(.)\). The activated output can be represented as shown in Equation (\ref{eq:7}).

\begin{equation} \label{eq:7}
    a_{i,j,k}^l= g(Z_{i,j,k}^l )    
\end{equation}

Additionally, there are pooling layers that introduce shift-invariance by reducing the resolution of the activated feature maps. Each pooling layer connects the feature map to the preceding convolutional layer. The expression for pooling is as shown in Equation (\ref{eq:8}).

\begin{equation} \label{eq:8}
    y_{i,j,k}^l = P(a_{n,m,k}^l ), \forall (m,n) \in R_{ij}
\end{equation}

Here \(P(.)\) is a pooling operation for the local neighborhood \(R_{ij}\) around the location \((i,j)\). In this work, we used CNN for both denoising as well as classification. The details of their architectures and their impacts are discussed in the result section.

\section{Results and Discussion}

\subsection{Performance of Denoising Algorithms}

For efficient and adaptive denoising, we analyzed various autoencoder schemes, starting with the fully-connected autoencoder. In our first iteration, we experimented with the fully connected network having three hidden layers with 512, 256, and 512 neurons, respectively. The input layer is the 1D vectorized array of the input cell diffraction pattern, e.g., of 66 × 66 pixels. The input to the model is the noisy version of the input cell diffraction pattern and the expected target output is the original cell diffraction pat-tern. The noisy cell diffraction patterns were created using a Gaussian distribution with variance ranging from 100 to 600 with zero mean (refer to the supplementary section (\ref{supplementary}) for detail). Further, we experimented with an increased network size having five hidden layers with 256, 128, 64, 128, and 256 neurons, respectively. In all these networks, rectified linear unit (ReLU) was used as the activation function while mean squared error (MSE) \cite{Ahmed2005,Zhang2018,Munir2019,Shanthi2019} was used to calculate the loss. The Adam optimizer \cite{Shen2019,Kingma2015} was found to deliver better convergence and hence used to perfect the weight and biases. The denoising performance was quantified in terms of the improvement in SNR, measured in dB, denoted here by $SNR_{imp}$, as given by Equation (\ref{eq:9}) \cite{Chiang2019}.

\begin{equation} \label{eq:9}
    {SNR}_{imp}  = {SNR}_{out}-{SNR}_{in}
\end{equation}

\begin{center}
    \[ where \qquad {SNR}_{out} = {10 \: log}_{10} \: \frac{ \sum_{n=1}^N x_i^2 }{ \sum_{n=1}^N (\hat{x_i} - x_i )^2 }  \qquad and \qquad  {SNR}_{in} = {10 \: log}_{10} \: \frac{ \sum_{n=1}^N x_i^2 }{ \sum_{n=1}^N (\tilde{x_i} - x_i )^2 } \]
\end{center}

Here $x_i$ is the value of sampling point $i$ in the original LSIT signal, $\tilde{x_i}$  is the value of sampling point $i$ in the noisy LSIT, and $\hat{x_i}$ is the value of sampling point \textit{i} in the denoised version of the same cell diffraction pattern. \(N\) is the total number of sample points in that LSIT image (cell diffraction patterns).

The fully connected network for both the above configuration shows no significant improvement in $SNR_{imp}$  after reaching saturation at around - 10.08 dB. For further improvement, we experimented with CNN architecture using various models with a different number of convolution layers and distinct kernel sizes. The configuration of the model which accomplished the best outcomes is 3×3, 3×3, 5×5, 5×5, 7×7, 7×7, 1×1 with 32 filters in each layer except the last layer. The last layer consists of a single pixel filter (1×1 filter) that is used to condense the output across all the 32 filters. Here, the input and output size are the same. Padding was used to maintain the original size after the output of each convolutional layer. Adam optimizer was used to optimize the network to reduce the mean squared error loss. The CNN results show a better reconstruction as shown in Figure \ref{fig:Fig2}.

\begin{figure}
  \centering
  \includegraphics[width=\textwidth,height=0.8\textheight,keepaspectratio]{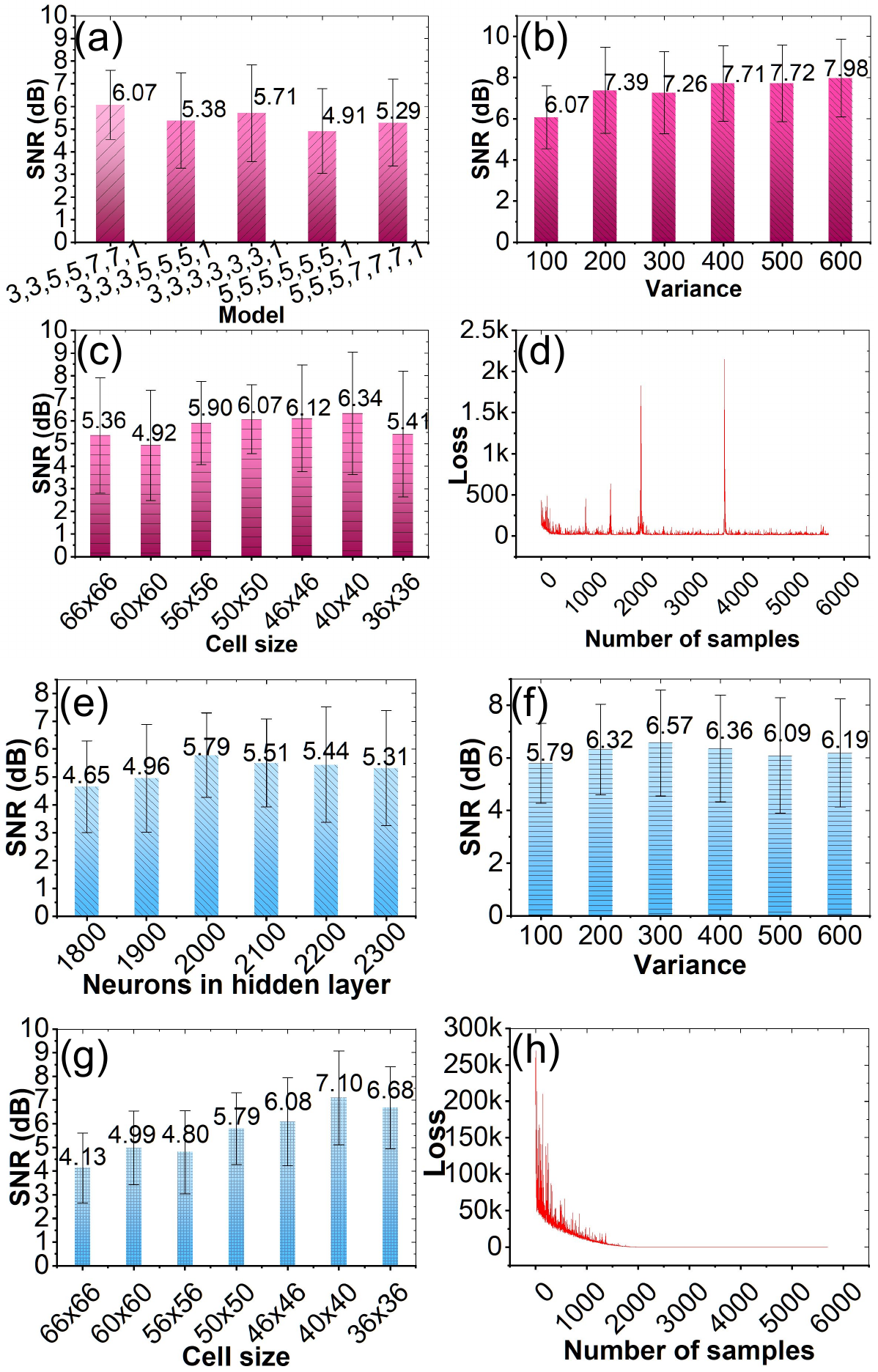}
  \caption{Results from the CNN and ELM autoencoder. (a) The performance of the CNN autoencoder for improved SNR (average of all the classes) across various layers and kernel sizes. (b) The performance of the CNN autoencoder across varying noise levels. Here, the variance of the Gaussian noise ranges from 100 to 600, and is evaluated on the optimal network architecture, i.e., 3,3,5,5,7,7,1. (c) The performance of the CNN autoencoder across various input sizes (cropping size). Here, the sizes vary from 66 × 66 to 36 × 36. (d) The convergence of the optimal CNN network with the number of samples for the first epoch. (e) The number of hidden layer neurons in ELM autoencoder vs. improvement in SNR, (f) Variance vs. improved SNR for ELM autoencoder, (g) Input size vs. improved SNR for ELM autoencoder, and (h) The convergence of ELM autoencoder within the 1st epoch.}
  \label{fig:Fig3}
\end{figure}

The CNN network has been optimized for various parameters. First, the optimization of the network for various design parameters, such as varying the convolution layers and the kernel sizes, was carried out. The results in Figure \ref{fig:Fig3}a show that the architecture with kernel sizes 3×3, 3×3, 5×5, 5×5, 7×7, 7×7, 1×1 has a better performance in terms of $SNR_{imp}$. The performance of the optimized network for various noise parameters, as shown in Figure \ref{fig:Fig3}b, indicates the network performs better reconstruction with increasing noise variance in the image (cell diffraction pattern). An increase in the variance results in a noisier image (cell diffraction pattern), which warrants a detailed reconstruction to reverse it to the original form, and hence larger the value of $SNR_{imp}$. Therefore, a higher improvement in $SNR_{imp}$ implies the network has learned the optimal representational features for the cell types which enables it to perform a better qualitative reconstruction. Fig. \ref{fig:Fig3}c compares the reconstruction performance of the model on different sizes of the input image (cell diffraction pattern). Due to the black-box nature of deep learning methods, we had to create datasets with multiple cell-signature dimensions, such that the smallest size just covered the central signature of the cell and increased the window size till it covered a significant background portion as well. The models were evaluated across varying cell sizes to determine the optimal signal to background ratio, the spatial extent up to which the models covered the features, and to study its effects on the model performance. This analysis is critical in understanding the model explainability and interpretability since having a size larger than the optimum increases the inclusion of background artifacts that affect denoising as well as overpower the cell signal while having a smaller one could exclude the important deterministic features of the cell signature. The convergence in the training phase of the network is as shown in Figure 3d. The results depict the loss across the first epoch, with a high variation in the initial phase, gets smoother towards the end of the first iteration. The advantage of this system is that it generalizes well for all cell lines using the same model.

Further, we tried the ELM architecture which is well known for its fast convergence \cite{Cambria2013}. The results in Figure \ref{fig:Fig3}e–h show the performance of the ELM architecture with varying number of neurons in the hidden layer. As it can be concluded from Figure \ref{fig:Fig3}e, the model with 2000 neurons provides better performance in terms of $SNR_{imp}$. Further, the optimized model has been used to test the performance across various noise levels as shown in Figure \ref{fig:Fig3}f. It is observed that the model maintains the $SNR_{imp}$ value on increasing the noise in the input image (cell diffraction pattern), i.e., the image (cell diffraction pattern) quality of the output relative to the input remains the same. The results of the model performance across different sizes of the input image (cell diffraction pattern), as shown in Figure \ref{fig:Fig3}g, indicate that the 40×40 is having a higher value of SNR. However, the variation is of 2 as compared to the variation for the size 50×50, that is of 1.5, which is the lowest compared to all the other sizes. Since CNN shows a substantial performance with lower variance for the 50×50 input size, therefore, we fixed it as optimal for all the further studies and comparisons. Figure \ref{fig:Fig3}h shows the loss across the first epoch for ELM which is remarkably high initially but converges faster, as compared to CNN, after training with only a few thousand samples. This faster convergence may help save time and resources during incremental training phases for newer cell types. The performances of these optimized models have been compared with the traditional denoising methods as shown in Table \ref{tab:table1} (The comparative visual reconstructions are provided in the supplementary document (Section \ref{supplementary})). It is concluded from the data that CNN shows better performance compared to the other modalities (The details of the traditional methods are provided in the supplementary document (Section \ref{supplementary})). Therefore, we prefer to use CNN for denoising.

\begin{table}
  \caption{Comparison of improved SNR with respect to the variance for various denoising modalities (here we keep input and output size as 50 × 50)}
  \centering
    \begin{tabular}{cccccccc}
    \toprule
    Variance & Gaussian & Average  & Median   & Bilateral & BM3D     & CNN     & ELM     \\
    \hline
    100 & 4.525580 & 4.237872 & 3.199475 & $-0.02635$  & 5.684597 & 6.06905 & 5.79161 \\
    200 & 4.613044 & 4.198564 & 3.068866 & $-0.00640$  & 5.713797 & 7.38544 & 6.31896 \\
    300 & 4.476552 & 4.259405 & 3.078091 & $-0.10943$  & 5.119646 & 7.26487 & 6.57256 \\
    400 & 4.674922 & 4.703331 & 3.405470 & $-0.05448$  & 3.792226 & 7.71265 & 6.36135 \\
    500 & 4.128021 & 4.201562 & 3.198964 & $-0.14542$  & 2.206876 & 7.72364 & 6.09254 \\
    600 & 4.023621 & 4.183404 & 2.908946 & $-0.16924$  & 1.683290 & 7.97877 & 6.19359 \\
    \bottomrule
  \end{tabular}
  \label{tab:table1}
\end{table}

\subsection{Performance of Classification Algorithm}

\begin{figure}
  \centering
  \includegraphics[width=\textwidth,keepaspectratio]{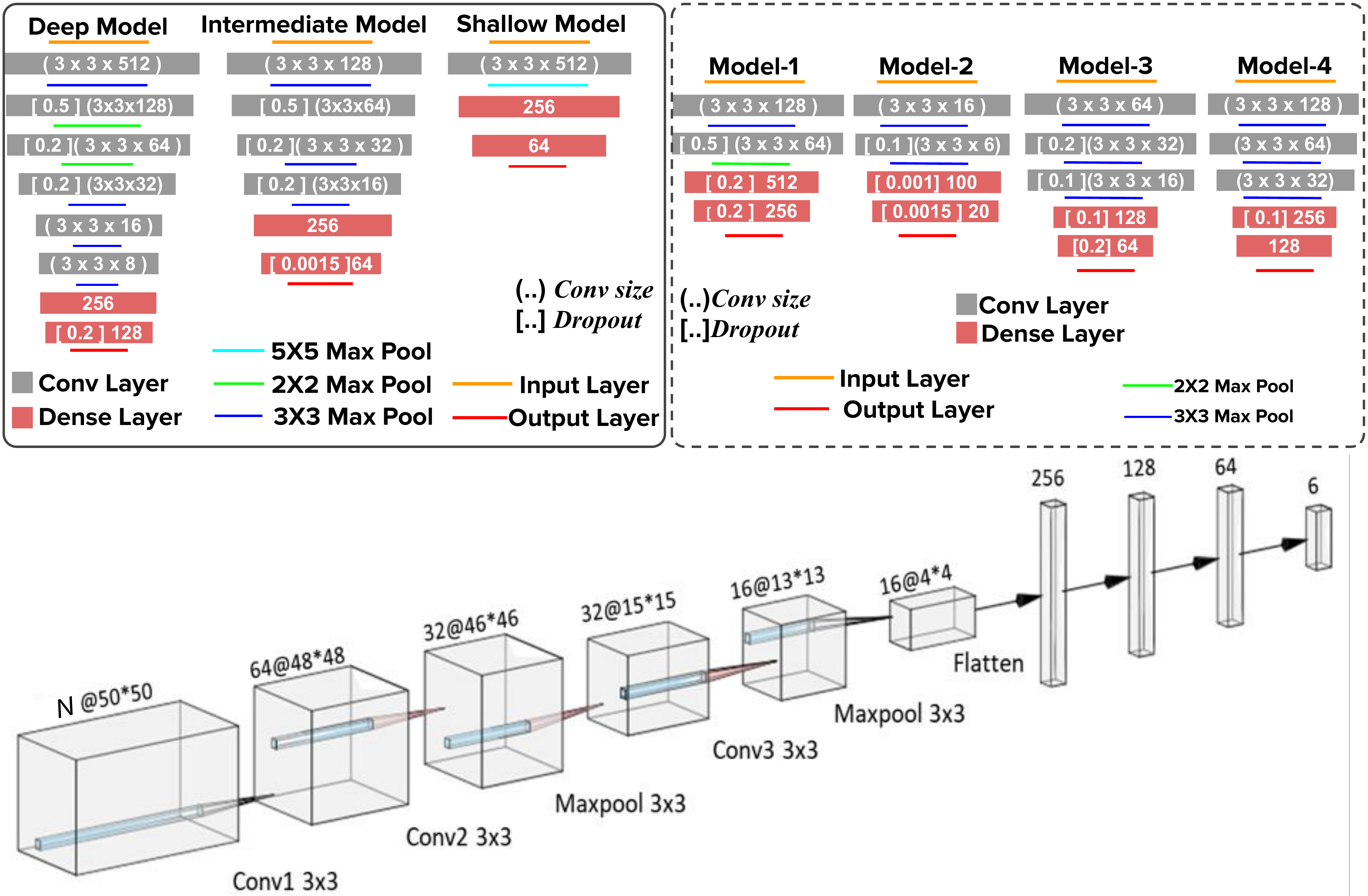}
  \caption{The CNN Architectures with varying depth. The models in the solid box are for the optimization of the model with varying depth. The models inside the dotted box are for the optimization of the parameters. Here, the orange line is the input layer and the rectangle represents the 2D convolution layer. The kernel size and number of kernels are indicated inside the round brackets, the dropout rate in square brackets, and the number of neurons in fully connected layers are placed directly within the rectangle. The aqua blue, green, and purple lines are the max-pool layers. The red line is the output layer and uses a SoftMax activation function. The final optimized architecture is as represented by the last 3D figure.}
  \label{fig:Fig4}
\end{figure}

Since the diffraction patterns of cells and microparticles in a LSIT micrograph depend upon their physical and optical properties, therefore, the diffraction patterns carry the unique signatures of each of the cell types as shown in the 2D contour plot in Figure \ref{fig:Fig2}. These unique signatures can be utilized for the classification of these cell types. Since our previous inference concludes that CNN works better for denoising, therefore we experimented with the same modality for the classification as well. In this work, in order to determine the optimal architecture of CNN for cell-line recognition, we first proceeded to find the optimal depth of the network by studying the classification performance of the model on increasing the depth, by adding convolutional and pooling layers, as well as by varying number of kernels and kernel size, till we reached performance saturation. We have experimented with and evaluated various shallow and deep CNN models to classify cell lines. The details of the model architecture are as described in Figure \ref{fig:Fig4}.

\begin{figure}
  \centering
  \includegraphics[width=\textwidth,keepaspectratio]{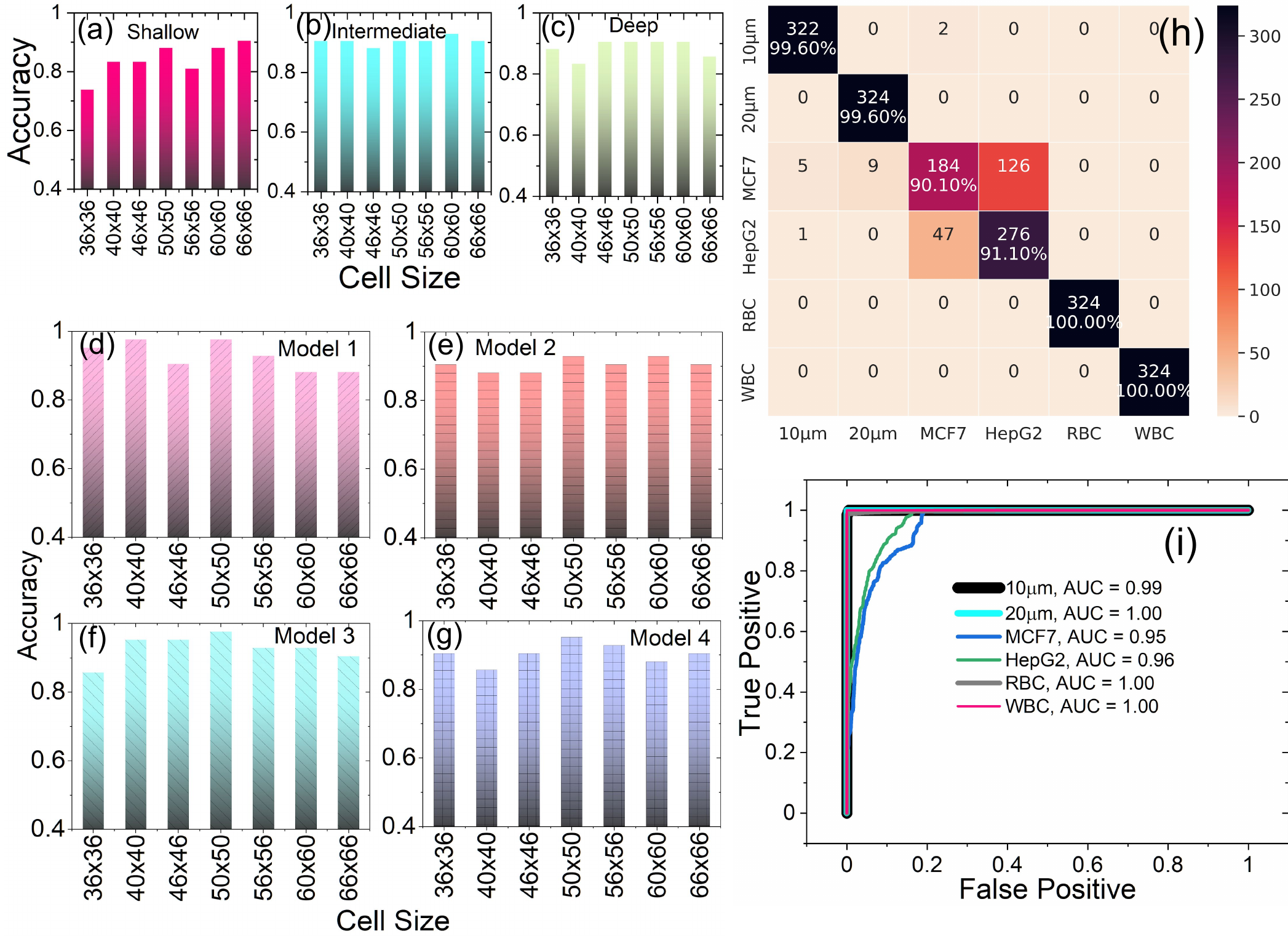}
  \caption{Results for the optimization of the CNN model. (a) Performance of the fine-tuned intermediate architecture. (b) The confusion matrix showing classification accuracy on the test dataset across all cell lines using the fine-tuned and optimized model. (c) The receiver operating characteristic (ROC) curve for each of the cell lines for the optimized model.}
  \label{fig:Fig5}
\end{figure}

The Deep Model starts with a convolutional (Conv2D) layer having 512 kernels of size 3×3, followed by a max-pooling layer of same kernel size. The output from this is further convoluted with 128 kernels of 3×3 size with a dropout rate of 0.5, and then a max pool with 2×2 kernel. This output goes to a Conv2D layer with 64 of 3×3 sized kernels and a dropout rate of 0.2. We further reduce the dimension using a 2×2 max pool kernel. The output of this layer further convolves with 32 of 3×3 kernels, then a dropout of 0.2. The output dimension from this convoluted layer is further reduced by using the max pool with a 3×3 kernel. This again convolves with 16 of 3×3 kernels, and a max pool layer with 3×3 kernel. The output of which is again convoluted with 8 of 3×3 kernels, followed by a 3×3 max pool. This output is then vectorized and input to a fully connected (FC) layer having 256 nodes, and then to another FC with 128 nodes and having a dropout of 0.2. The final layer is a SoftMax function, with 6 output nodes. The model architectures used for studying the impact of the network depth and breadth on performance are well described in Figure \ref{fig:Fig4}. Once the approximate optimal depth and breadth had been determined, we proceeded to fine-tune the hyper-parameters such as the number of kernels, kernel size, and dropouts, to reach the best performance of the models across varying cell sizes, i.e., input dimension of cells. In all the models, Adam \cite{Wang2019} provided better convergence as compared to other optimizers and has been used as the model optimizer, with categorical cross-entropy \cite{Zhang2020,Raczkowski2019} as a loss estimator. The result for depth and breadth optimization indicates that the average accuracy of the intermediate model is >0.85 (including all the classes), whereas it is <0.85 for shallow and deep networks. Therefore, we proceed with this intermediate model as an optimum model for further study. Considering the intermediate model has the optimum breadth and depth, the further optimization of the parameters and the results are as shown in Figure \ref{fig:Fig5}a.

From Figure \ref{fig:Fig5}a, it is inferred that Model 3 shows better classification performance, on the validation fold, of all models. The results depict that there is consistency in performance for the input sizes 40×40 to 66×66, with a very small variance in the accuracy. The performance of this optimized model is further evaluated over the test dataset containing 324 samples of each cell type. The per-class performance of this model is shown in the confusion matrix of Figure \ref{fig:Fig5}b. The results depict that the model can classify RBC, WBC, 10 µm, and 20 µm bead with over 99\% accuracy. However, the comparatively poor performance of about 90\% for the cancer cells, HepG2 and MCF7, can be attributed to the non-homogeneity in their signature characteristics as well as the lack of sufficient original samples which further complicates the issue. This is well depicted in our previous work \cite{Roy2016} (see Figure \ref{fig:Fig2} of the reference). From the receiver operating characteristic (ROC) curve for all the cell lines shown in Figure \ref{fig:Fig5}c, the area under the curve (AUC) for all the cell lines is $>$0.99, except MCF7 ($\sim$0.95) and HepG2 ($\sim$0.96). From these results, it can be inferred that the classifier is working well, especially for RBC, WBC, 10 µm, and 20 µm beads. The visualization of the internal activation maps, as shown in the supplementary (Section \ref{supplementary}), implies that the network is learning core descriptive features in the diffraction signatures rather than using some random feature.

The performance evaluation of the proposed Al model with various matrices such as true positive (TP), true negative (TN), false positive (FP), false negative (FN), accuracy, recall, specificity, sensitivity, F1 score, positive predictive value (PPV) and negative predictive value (NPV) is summarized in Table \ref{tab:table2}.

\begin{table}
  \caption{Performance evaluation scores of the proposed AI model using various metrics for different cell types}
  \centering
  \begin{adjustbox}{width=\textwidth}
    \begin{tabular}{ccccccccccccc}
    \toprule
    Sample Type & TP  & TN   & FP  & FN  & Accuracy & Precision & Recall & Specificity & Sensitivity & F1     & PPV    & NPV    \\
    \hline
    $10\mu m$ bead  & 322 & 1614 & 2   & 6   & 0.9959   & 0.9938    & 0.9817 & 0.9988      & 0.9817      & 0.9877 & 0.9938 & 0.9963 \\
    $20\mu m$ bead & 324 & 1611 & 0   & 9   & 0.9954   & 1.0000    & 0.9730 & 1.0000      & 0.9730      & 0.9863 & 1.0000 & 0.9944 \\
    MCF7        & 184 & 1571 & 140 & 49  & 0.9028   & 0.5679    & 0.7897 & 0.9182      & 0.7897      & 0.6607 & 0.5679 & 0.9698 \\
    HepG2       & 276 & 1494 & 48  & 126 & 0.9105   & 0.8519    & 0.6866 & 0.9689      & 0.6866      & 0.7603 & 0.8519 & 0.9222 \\
    RBC         & 324 & 1620 & 0   & 0   & 1.0000   & 1.0000    & 1.0000 & 1.0000      & 1.0000      & 1.0000 & 1.0000 & 1.0000 \\
    WBC         & 324 & 1620 & 0   & 0   & 1.0000   & 1.0000    & 1.0000 & 1.0000      & 1.0000      & 1.0000 & 1.0000 & 1.0000 \\
    \bottomrule
  \end{tabular}
  \end{adjustbox}
  \label{tab:table2}
\end{table}

\begin{figure}
  \centering
  \includegraphics[width=\textwidth,keepaspectratio]{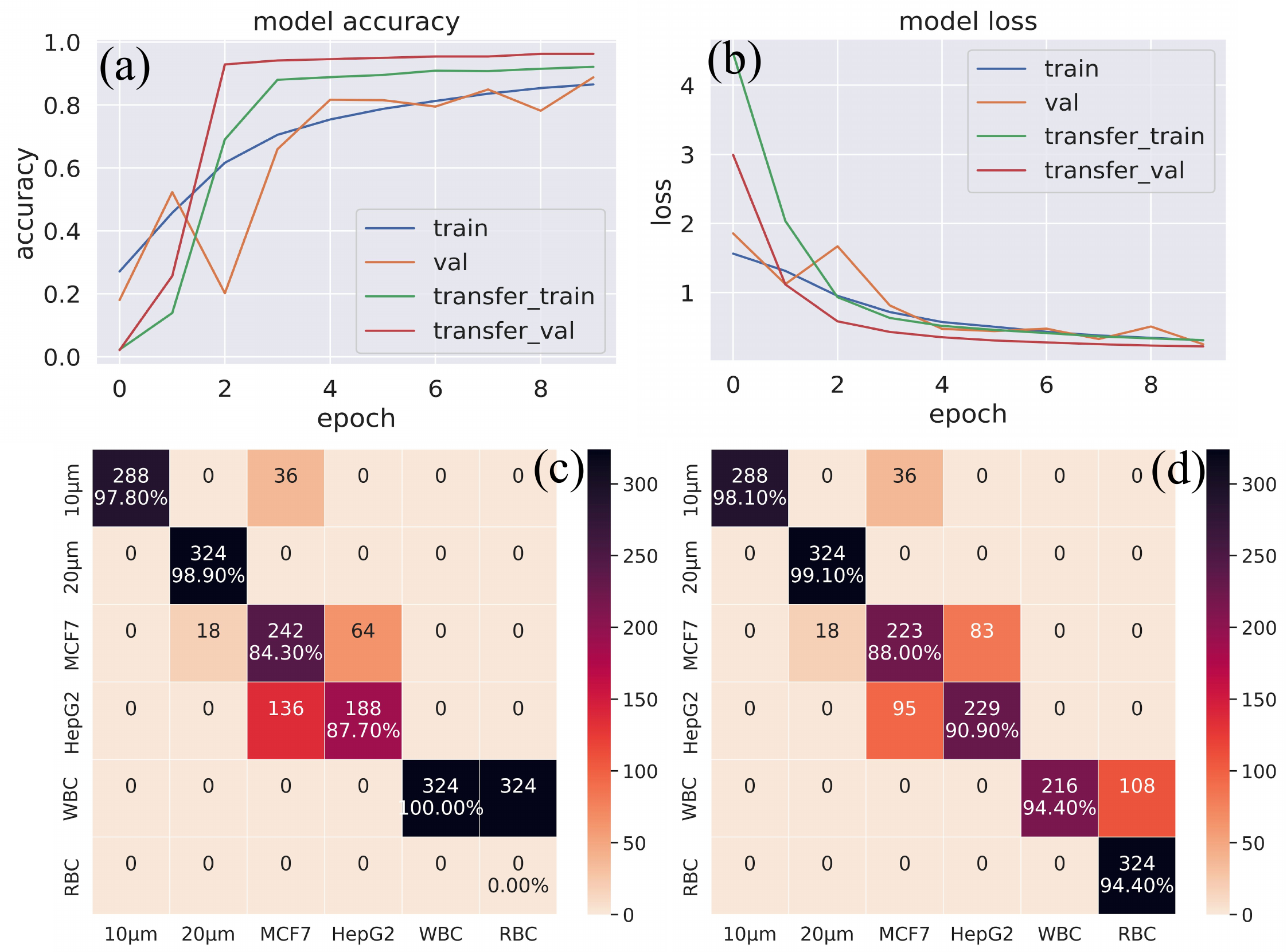}
  \caption{The results from the transfer of learning study. (a) epoch vs. accuracy graph for the initial epochs in the initial training phase without RBC and then the transfer of training with RBC. The blue and saffron colors represent the initial training accuracy and validation curves. The green and red lines represent the transfer learning accuracy and validation curves, (b) epoch vs. loss, (c) the confusion matrix from the pre-trained model, (d) the confusion matrix with the re-trained model.}
  \label{fig:Fig6}
\end{figure}

Additionally, we also investigate the transfer of learning to gauge the ability of the trained network to adapt to newer cell types (Figure \ref{fig:Fig6}). For this scenario, the CNN was initially trained with all the cell lines except RBC. From the epoch versus accuracy graph in Figure \ref{fig:Fig6}a, the transfer training achieved higher accuracy with the same number of epochs compared to the initial training. This is also validated by the epoch versus loss graph in Figure \ref{fig:Fig6}b. From these results, it can be inferred that the network can be effectively used to adapt to newer cell lines with very less amount of training. From the per-class test accuracy shown in Figure \ref{fig:Fig6}c, it is observed that the model misclassified all the RBC samples as WBC. In the transfer of the learning phase, the initially trained network is frozen except for the last layer, which is modified to accommodate the newer class and kept trainable. The network is then re-trained with a mix of RBC samples. The per-class test accuracy of the re-trained model is shown in the confusion matrix of Figure \ref{fig:Fig6}d, where it is inferred that the re-trained network can classify RBC correctly with substantial accuracy.

\begin{figure}
  \centering
  \includegraphics[width=\textwidth,keepaspectratio]{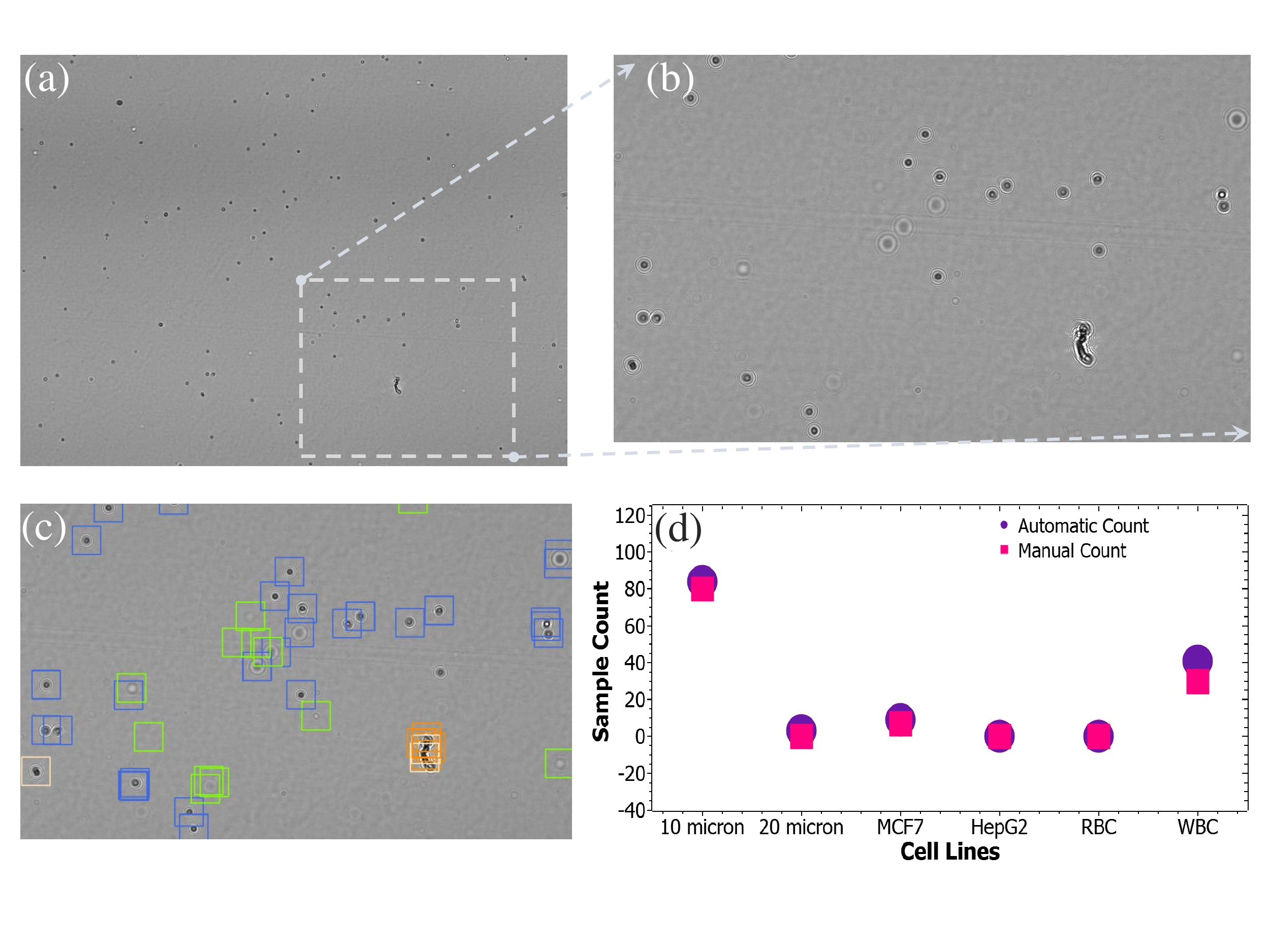}
  \caption{Comparison between the proposed AI method and the manual method for counting. (a) A whole frame LSIT micrograph, (b) magnified region-of-interest (ROI) of the whole frame LSIT micrograph, (c) automatically classified cell diffraction patterns of the ROI, and (d) comparison of the cell counts between those two different modalities.}
  \label{fig:Fig7}
\end{figure}

The comparison between the proposed AI method and the manual method for counting various cell types from a heterogeneous sample is shown in Figure \ref{fig:Fig7}. This comparison shows the robustness of the model.

\section{Conclusion}

In conclusion, we have explored the advantage of neural networks in the characterization of LSIT micrographs. Here, we have perfected neural networks that can automatically improve the signal quality and classify the cell types. We find that this neural network can classify the RBC, WBC with great accuracy, i.e., over 98\%, and the cancer cell with an accuracy of about 90\%. This network is also flexible for adapting to newer cell lines by retraining the trained network with very few samples. Together with this algorithm, the lightweight and cost-effective LSIT setup can be utilized as a point of care system for the diagnosis of pathological disorders in the resource-limited setup of our world. In our future work, we aim to combine the denoising and classification modalities due to the significant overlap in their operation. This will remove the dual training times as well as minimize computation costs. Also, we aim to work on improving their performance and deploying it in real scenarios.

\section*{Acknowledgments}

\textbf{Funding:} M.R. acknowledges the seed grant No. ORSP/R\&D/PDPU/2019/MR/RO051 of PDPU (the AI software development part) and the core research grant No. CRG/2020/000869 of the Science and Engineering Research Board (SERB), India. S.S. acknowledges the support of the Basic Science Research Program (Grant\#: 2014R1A6A1030732, Grant\#: 2020R1A2C1012109) through the National Research Foundation (NRF) of Korea, the Korea Medical Device Development Fund (Project\#: 202012E04) granted by the Korean government (the Ministry of Science and ICT, the Ministry of Trade, Industry and Energy, the Ministry of Health \& Welfare, the Ministry of Food and Drug Safety), and the project titled ‘Development of Management Technology for HNS Accident’ and ‘Development of Technology for Impact Assessment and Management of HNS discharged from Marine Industrial Facilities’, funded by the Ministry of Oceans and Fisheries of Korea.

\textbf{Institutional Review Board Statement:} All human blood samples were approved by the Institutional Review Board of Anam Hospital, Korea University (\# 2021AN0040).

\bibliographystyle{unsrt}  

\vspace{10mm}

\section{Supplementary Information} \label{supplementary}

\subsection{Convolutional Neural Network (CNN) workflow}

\begin{table}[h]
\centering
\begin{tabular}{llll}
Network Architecture &   & \qquad Output dimensions   \\
0-Input              & : & N, 50x50 (batch size, input image)   \\
1-Conv2D             & : & 64, 3x3, \qquad \qquad 48x48, Relu \\
2-Conv2D             & : & 32, 3x3, \qquad \qquad 46x46, Relu \\
3-MaxPool2D          & : & 32, 3x3, \qquad \qquad 15x15       \\
4-Conv2D             & : & 16, 3x3, \qquad \qquad 13x13, Relu \\
5-MaxPool2D          & : & 16, 3x3, \qquad \qquad 4x4         \\
6-Flatten            & : & 16, 4x4, \qquad \qquad 256         \\
7-FC                 & : & \: \quad \qquad \qquad \qquad 128, Relu   \\
8-FC                 & : & \: \quad \qquad \qquad \qquad 64, Relu    \\
9-FC                 & : & \: \quad \qquad \qquad \qquad 6, Softmax 
\end{tabular}
\end{table}

Here, FC represents Fully Connected layers in the CNN architecture. \\

\textbf{Forward Propagation}

\underline{Step1:} \\

Given an input image $I$ (50x50 single grayscale image) to be convolved with $\prime p \prime$ kernels, the 2D convolution operation C is defined as

\begin{center}
    \( C_p^1=f( I*k_{1,p}^1  +b_p^1  ) \) for p=1 to 64
\end{center}

where for $k_{1,p}^1$ : “1” in the superscript denotes the hierarchical order of the convolution operation and $(1, p)$ in the subscript denotes 1 input channel, $p^{th}$ output channel

\qquad \qquad $b_p^1$ represents the bias corresponding to each of the p kernels

\qquad \qquad $*$ represents the convolution (element-wise product followed by addition) between the input and the kernel

\qquad \qquad $f$ represents Relu activation, applied to the output of the convolution operation\\

\underline{Step2 $:$}\\
\begin{center}
    \( C_q^2 = f( \Sigma_p C_p^1 *k_{p,q}^2  +b_q^2  ) \) for p = 1 to 64; q= 1 to 32 \\
\end{center}

\underline{Step3 $:$}\\
\begin{center}
    \( M_q^1 (i,j) = max( C_q^2 [ 3(i-1)+1 : 3(i-1)+3 ,3(j-1)+1 : 3(j-1)+3 ]) \) for i,j = 1 to 1; q= 1 to 32;     \\
\end{center}

\underline{Step4 $:$}\\
\begin{center}
    \( C_r^3 = f( \Sigma_q M_q^1 *k_{q,r}^3  +b_r^3  )\) for q= 1 to 32; r = 1 to 16 \\
\end{center}

\underline{Step5 $:$}\\
\begin{center}
    \( M_r^2 (i,j) = max( C_r^3 [ 3(i-1) + 1 : 3(i-1) + 3, 3(j-1) +1 : 3(j-1) + 3 ]) \) for i,j = 1 to 4; r = 1 to 16;\\
\end{center}

\underline{Step6 $:$}\\
\begin{center}
    \(g =G( \{M_r^2\} \)  ;r=1 to 4  );   (flatten/vectorization)
\end{center}

The rest of the steps for the fully connected layers are the same as in Feedforward Multi-Layer Perceptron (MLP). \\

\textbf{Backward Propagation}\\

Let L be the loss at the end of the epoch during the training phase.

All the equations for backpropagation for the fully connected layers are the same as in MLP till the FC-7 layer. \\

\underline{Step1:}

Those computed gradients from layer FC-7 simply need to be vectorized into matrix 4x4x16.

Reshape output error vector $\Delta g$ (256x1) into (4x4x16):

\begin{center}
    \( \{\Delta M_r^2\}_{r=1 to 4} = G^{-1} ( \Delta g ); \) (flatten/vectorization)    
\end{center}

Since it is a max-pool operation, it does not involve any parameters, hence the stored mask in the forward pass is used as a reference for the backpropagation update path, giving $\Delta C_{r}^3$. \\

\underline{Step2: (Inverse of step 4 in the forward path)}

Calculating $\Delta k_{q,r}^3$ :

\begin{center}
 \[ \Delta k_{q,r}^3 (u,v)= \frac{\partial L}{\partial k_{q,r}^3 (u,v)} = \sum_i^{13} \sum_j^{13} \frac{\partial L}{\partial C_{r}^3 (i,j)} \frac{\partial C_{r}^3 (i,j)}{\partial k_{q,r}^3 (u,v)} \]
 
 \[ = \sum_i^{13} \sum _j^{13} \Delta C_{r}^3 (i,j) . \frac{\partial C_{r}^3 (i,j)}{\partial t} \frac{\partial t}{\partial k_{q,r}^3 (u,v) } ; \quad t = ( \sum_q M_q^1 *k_{q,r}^3 + b_r^3) \]

 \[ =\sum _i^{13} \sum_j^{13} \Delta C_{r}^3 (i,j) . f'. \frac{\partial  (\sum_q M_q^1 *k_{q,r}^3 + b_r^3)}{\partial k_{q,r}^3 (u,v)} \]

 \[ =\sum _i^{13} \sum_j^{13} \Delta C_{r}^3 (i,j) .f'. M_q^1  (i-u,j-v) \]  
 
 \[ Let \: \Delta C_{r,f}^3 (i,j) = \Delta C_{r}^3 (i,j) .f' \]
\end{center}

Hence, we get a deconvolution as follows:

\begin{center}
    \[ \Delta k_{q,r}^3 (u,v) = \sum_i^{13} \sum_j^{13} M_q^1  (i-u,j-v) . \Delta C_{r,f}^3 (i,j) \]
\end{center}

Here, $M_q^{1^{inv}}$ which is equivalent to convolving with the inverse of $M_q^1$ i.e. rotating $M_q^1$ by 180º (flipped)

\begin{center}
    \[ \Delta k_{q,}^3=  M_q^{1^{inv}}  * \Delta C_{r,f}^3 \]    
\end{center}

Calculating $\Delta M_q^1$:

Similarly, following the above process for $\Delta M_q^1$, we get similar deconvolution:

\begin{center}
    \[ C_r^3 = f ( \sum_q M_q^1 *k_{q,r}^3 + b_r^3 ); \]
\[ \Delta M_q^1 (i,j) = \frac{\partial L}{\partial M_q^1 (i,j)} \]
\[ \Delta M_q^1 = \sum_r^{15} \Delta C_{r,f}^3 * k_{q,r}^{3^{inv}}  \]
\end{center}

Calculating $\Delta b_r^3$:

\begin{center}
    \[ \Delta b_r^3 = \frac{\partial L}{\partial b_r^3} = \sum_i^{13} \sum_j^{13} \frac{\partial L}{\partial C_(r )^3 (i,j)} \frac{\partial C_(r )^3 (i,j)}{\partial b_r^3} \]
    
    \[ =\sum_i^{13} \sum_j^{13} \Delta C_{r}^3 (i,j) .\frac{\partial C_{r}^3 (i,j)}{\partial t } \frac{\partial t}{\partial b_r^3 }; \qquad t=( \sum_q M_q^1 * k_{q,r}^3 + b_r^3) \]
    
    \[ = \sum_i^{13} \sum_j^{13} \Delta C_{r}^3 (i,j) . f' . \frac{\partial (\sum_q M_q^1 * k_{q,r}^3 + b_r^3 )}{\partial b_r^3} \]
    
    \[ = \sum_i^{13} \sum_j^{13} \Delta C_{r}^3 (i,j) . f' \]
    
    \[ \Delta b_r^3 = \sum_i^{13} \sum_j^{13} \Delta C_{r,f}^3 (i,j) \] \\
\end{center}

\underline{Step3: (Inverse of step 3 in the forward path)}\\

Since it is a max-pool operation, it does not involve any parameters, the stored mask in the forward pass is used as a reference for backpropagation. Updates from $\Delta M_q^1$ will be propagated into respective $\Delta C_q^2$. \\

\underline{Step4: (Inverse of step 2 in the forward path)} \\

Calculating  $\Delta k_{p,q}^2$:

\begin{center}
    \[ \Delta k_{p,q}^2 (u,v) = \frac{\partial L}{\partial k_{p,q}^2 (u,v)} = \sum_i^{46} \sum_j^{46} \frac{\partial L}{\partial C_{q}^2 (i,j)} \frac{\partial C_{q}^2 (i,j)}{\partial k_{p,q}^2 (u,v)}  \]   
    
    \[ = \sum_i^{46} \sum_j^{46} \Delta C_{q}^2 (i,j) . \frac{\partial C_{q}^2 (i,j)}{\partial t} \frac{\partial t}{\partial k_{p,q}^2 (u,v)}; \qquad t= ( \sum_p C_p^1 * k_{p,q}^2 + b_q^2 ) \]
    
    \[ = \sum_i^{46} \sum_j^{46} \Delta C_{q}^2 (i,j) .f'. \frac{\partial ( \sum_p C_p^1 * k_{p,q}^2 + b_q^2) }{\partial k_{p,q}^2 (u,v) } \]
    
    Equivalently,

    \[ \Delta k_{p,q}^2=  C_p^{1^{inv}} * \Delta C_{q,f}^2 \]
\end{center}

Calculating $\Delta C_p^1$:

Similarly, following the above process for $\Delta M_q^1$, we get similar deconvolution:

\begin{center}
    \[ \Delta C_p^1 (i,j) = \frac{\partial L}{\partial C_p^1 (i,j)} \]

    \[ \Delta C_q^1= \sum_q^{48} \Delta C_{q,f}^2 * k_{p,q}^{2^{inv}} \]
\end{center}
  
Calculating $\Delta b_q^2$:

\begin{center}
    \[ \Delta b_q^2 = \frac{\partial L}{\partial b_q^2} = \sum_i^{46}  \sum_j^{46} \frac{\partial L}{\partial C_{q}^2 (i,j)} \frac{\partial C_{q}^2 (i,j)}{\partial b_q^2} \]

    \[ \Delta b_q^2 = \sum_i^{46} \sum_j^{46} \Delta C_{q,f}^2 (i,j) \]  \\
\end{center}

\underline{Step5: (Inverse Of step 1 in the forward path)}\\

Calculating  $\Delta k_{1,p}^1$ :

\begin{center}
    \[ \Delta k_{1,p}^1 (u,v) = \frac{\partial L}{\partial k_{1,p}^1 (u,v) } = \sum_i^{50} \sum_j^{50} \frac{\partial L}{\partial C_{p}^1 (i,j)} \frac{\partial C_{p}^1 (i,j)}{\partial k_{1,p}^1 (u,v)} \]   
    
    Equivalently,
    
    \[ \Delta k_{1,p}^1= I^{inv} * \Delta C_{p,f}^1 \]
\end{center}
 
Calculating $\Delta b_p^1$:

\begin{center}
    \[ \Delta b_p^1 = \frac{\partial L}{\partial b_p^1} = \sum_i^{50} \sum_j^{50} \frac{\partial L}{\partial C_{p}^1 (i,j) } \frac{\partial C_{p}^1 (i,j)}{\partial b_p^1}   \]
    
    \[ \Delta b_p^1= \sum_i^{50} \sum_j^{50} \Delta C_{p,f}^1 (i,j)  \] \\
\end{center}

\vspace{2mm}

\subsection{Gaussian Noise}

The probability density function $p(x)$ for a Gaussian distribution is:

\begin{center}
    \[ p(x) = \frac{1}{\sqrt{2 \pi \sigma^2}} e^\frac{ - {(x - \mu)}^2 }{2 \sigma^2}\]     
\end{center}

where $\mu$ is the mean

\qquad \quad $\sigma$ is the standard deviation

\qquad \quad $\sigma^2$ is the variance

The generated noise is then added onto the image as

\begin{center}
    \[ \tilde{z} = z + n \]
\end{center}

where $z$ is the noise-free image

\qquad \quad $n$ is the noise with Gaussian distribution with desired variance (noise) \\


\subsection{Traditional denoising methods}

\textbf{Gaussian filtering:}

In this, the source image is convolved with the selected Gaussian filter for denoising.

The Gaussian distribution in 1-D has the form:
\begin{center}
    \[ G(x) = \frac{1}{\sqrt{2 \pi \sigma^2}} e^\frac{ - {x}^2 }{2 \sigma^2} \]        
\end{center}

where $\sigma$ is the standard deviation of the distribution. (Here, it is assumed that the distribution has a mean of zero)

Library used – OpenCV (cv2)

Parameters used – kernel size – 3*3 (standard)

Link to the library - \url{https://docs.opencv.org/master/d4/d86/group__imgproc__filter.html#gaabe8c836e97159a9193fb0b11ac52cf1} \\

\textbf{Average filtering:}

Average (or mean) filtering is a method of ‘smoothing’ images by reducing the amount of intensity variation between neighbouring pixels. The average filter works by moving through the image pixel by pixel, replacing each value with the average value of neighbouring pixels, including itself.

The function smooths an image using the kernel:

\begin{center}
    \[ K = \frac{1}{kernel_{height} * kernel_{width}} 
    \begin{bmatrix}
    1 & \cdots & 1 \\
   \vdots & \ddots & \vdots \\
   1 & \cdots & 1\\
\end{bmatrix} \]    
\end{center}

Library used – OpenCV (cv2)

Parameters used – kernel size – 3*3 

Link to the library - \url{https://docs.opencv.org/master/d4/d86/group__imgproc__filter.html#ga8c45db9afe636703801b0b2e440fce37} \\

\textbf{Median filtering:}

Median filtering is a non-linear method used to remove noise from images. It is widely used as it is very effective at removing noise while preserving edges. It is particularly effective at removing ‘salt and pepper’ type noise. The image is smoothened using the median filter with the kernel size × kernel size aperture. Each channel of a multi-channel image is processed independently.

Library used – OpenCV (cv2)

Parameters used – kernel size – 3*3 

Link to library - \url{https://docs.opencv.org/master/d4/d86/group__imgproc__filter.html#ga564869aa33e58769b4469101aac458f9} \\

\textbf{Bilateral filtering:}

The basic idea underlying bilateral filtering is to do in the range of an image that traditional filters do in its domain. Two pixels can be close to one another, that is, occupy the nearby spatial location, or they can be similar to one another, that is, have nearby values, possibly in a perceptually meaningful fashion. Bilateral Filter can reduce unwanted noise very well while keeping edges fairly sharp. 

Library used – OpenCV (cv2)

Parameters used – $kernel size = 3$, $sigmaColor=3$, $sigmaSpace=3$

Link to the library - \url{https://docs.opencv.org/master/d4/d86/group__imgproc__filter.html#ga9d7064d478c95d60003cf839430737ed} \\

\textbf{bm3d filtering:}

bm3d is based on an enhanced sparse representation in the transform-domain. The enhancement of the sparsity is achieved by grouping similar 2D image fragments (e.g. blocks) into 3D data arrays which we call "groups".

Image fragments are grouped based on similarity, but unlike standard k-means clustering and such cluster analysis methods, the image fragments are not necessarily disjoint. This block-matching algorithm is less computationally demanding. 

Library used – bm3d

Parameters used – \textit{sigma\_psd} = 15

Link to library - \url{https://pypi.org/project/bm3d/};\qquad \url{http://www.cs.tut.fi/~foi/GCF-BM3D/} \\

\subsection{Comparison of the denoised outputs from various modalities}

\begin{figure}[h!]
  \centering
  \includegraphics[width=\textwidth,keepaspectratio]{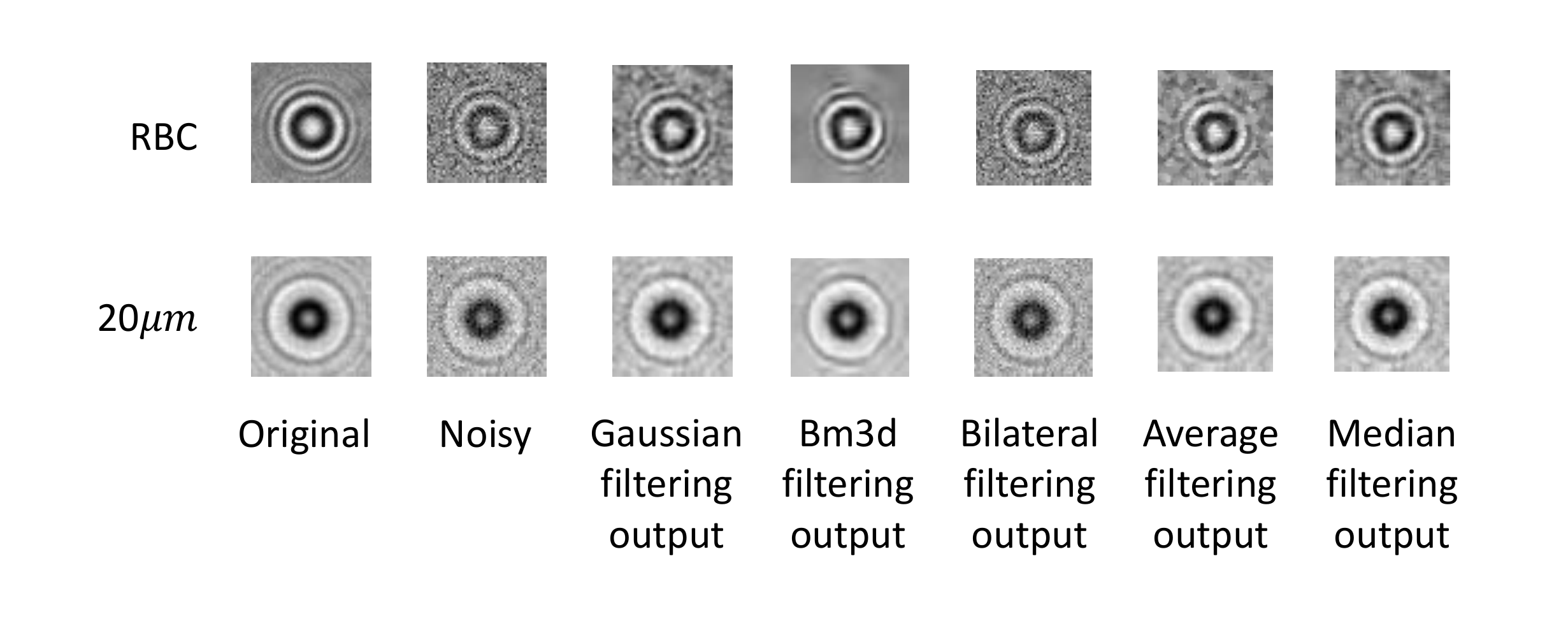}
  \label{fig:Fig8}
\end{figure}

\subsection{SNR of individual samples of various cell types}

\begin{figure}[h!]
  \centering
  \includegraphics[width=\textwidth,keepaspectratio]{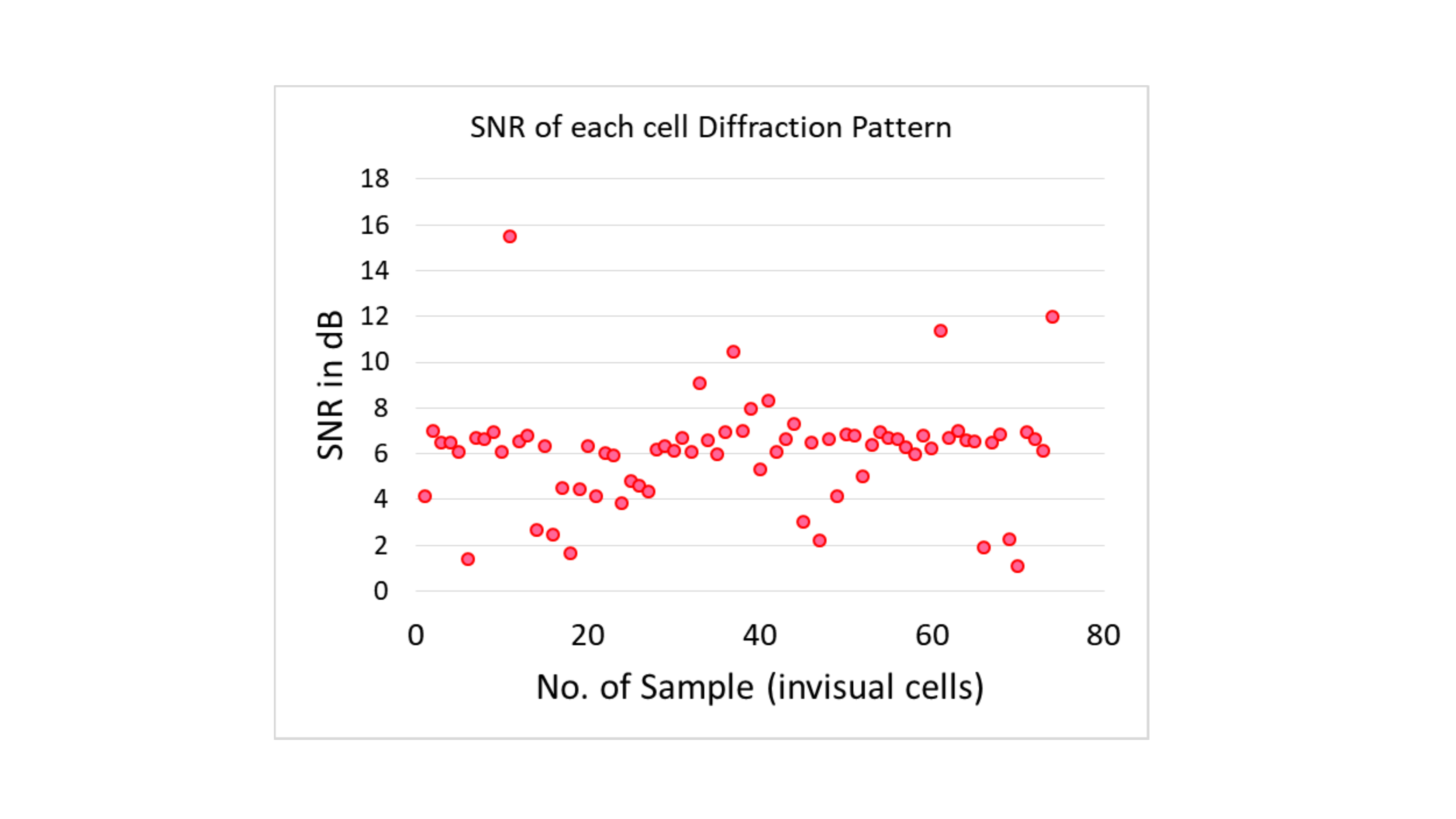}
  \caption{Graph depicting the variation in SNR of individual samples of various cell types denoised by the CNN Autoencoder}
  \label{fig:Fig9}
\end{figure}

\subsection{Grad-CAM and Saliency maps of the individual cell-lines}

\begin{figure}[h!]
  \centering
  \includegraphics[width=\textwidth,keepaspectratio]{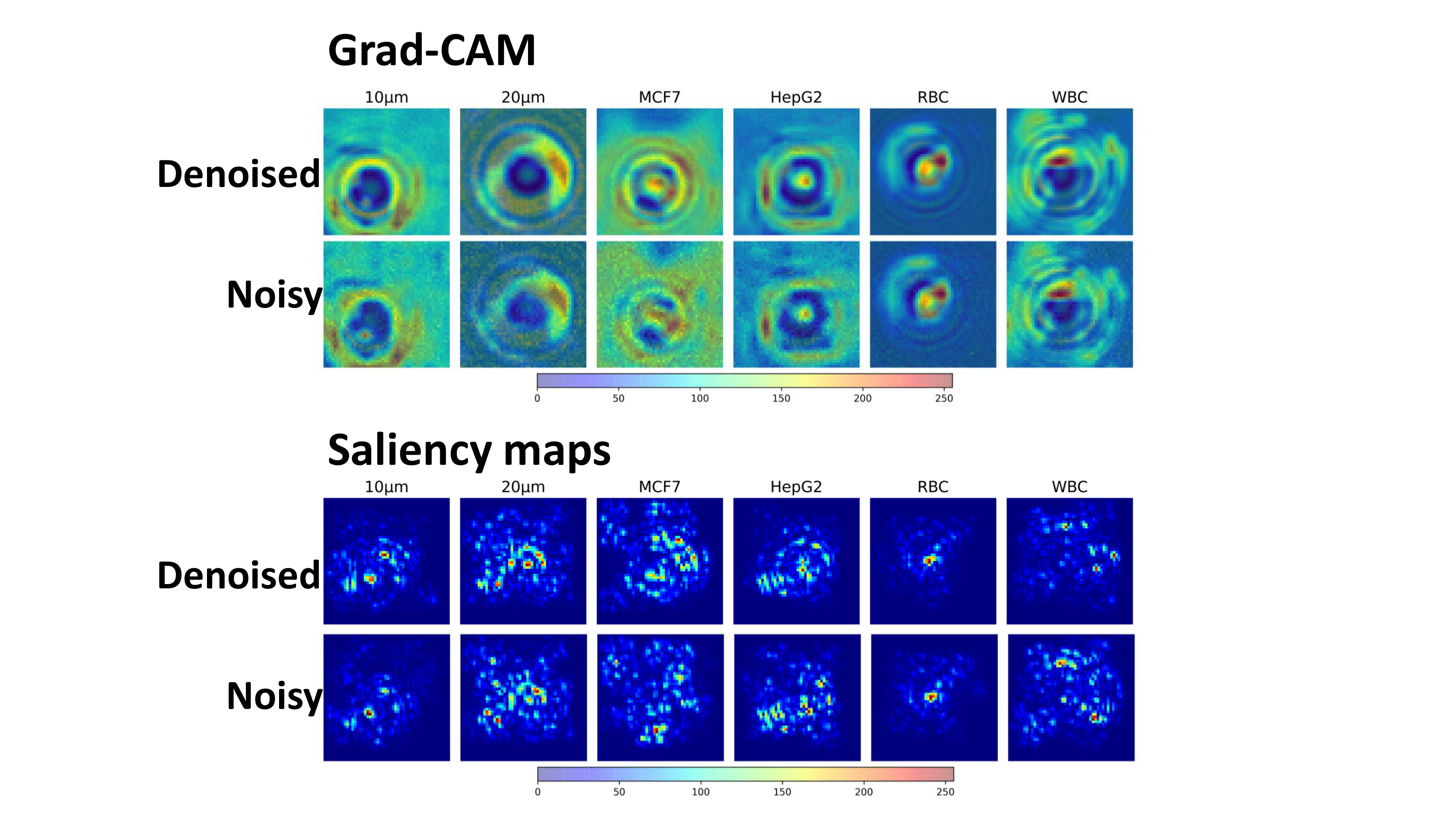}
  \label{fig:Fig10}
\end{figure}

Grad-CAM is a technique to measure the cumulative activations from all the layers of the network recorded across the image. It helps to localize the key regions in the image that contribute to the classification decision and provides a visual understanding of the same through the overlaid heat maps. As referenced from the colormap, the regions in red are of higher importance compared to those in blue.
From a comparison of noisy and denoised samples, it is observed that the significant activations in the Grad-CAMs are recorded at the center and on the diffraction rings of the cell-lines. The noise in the images skews the activations by either diminishing them in regions of interest or by increasing activations outside those regions. This skew is however corrected by the denoising as shown in Fig. \ref{fig:Fig10}.

Again, Saliency Maps help to identify the differentiating regions of interest that have a significant role in determination of the class label of the images. From a comparison of saliency, it is observed that denoising suppresses the pseudo-important regions of interest that have been generated due to the noise in the image.

\end{document}